# Enhancing Accessibility in Soft Robotics: Exploring Magnet-Embedded Paper-Based Interactions

**Ruhan Yang**
ATLAS Institute
University of Colorado Boulder
ruhan.yang@colorado.edu

**Ellen Yi-Luen Do**
ATLAS Institute
University of Colorado Boulder
ellen.do@colorado.edu

## ABSTRACT
This paper explores the implementation of embedded magnets to enhance paper-based interactions. The integration of magnets in paper-based interactions simplifies the fabrication process, making it more accessible for building soft robotics systems. We discuss various interaction patterns achievable through this approach and highlight their potential applications.

## CCS Concepts
• Human-centered computing~Interaction design~Interaction design process and methods

## 1. INTRODUCTION

Paper, due to its flexibility and adaptability, plays a crucial role in the development of soft robotics. It is not only environmentally friendly but also readily available, making it a popular choice for prototyping. However, the development of paper-based interactions has faced certain challenges. Traditionally, paper-based interactions required complex folding techniques or electronic components, posing high entry barriers. We aim to address these limitations by democratizing the fabrication of paper-based interactions, thus lowering the barriers for researchers and hobbyists in the field of soft robotics and related domains.

By utilizing magnets as a means to enable mechanical movements on paper, the prototyping and fabrication process of paper-based interactions becomes simplified. This combination leverages both the flexibility of paper and the versatile movement provided by magnetism. The integration of magnets and paper can be easily achieved using tape, eliminating the need for specialized tools or materials. In this paper, we discuss paper interfaces built using this approach and their mechanisms.

## 2. RELATED WORK

This work draws inspiration from research on paper-based interactions and the utilization of magnets in tangible interaction designs. Previous studies have employed paper as a medium to construct electronic products and prototype interface designs, with paper circuit-based interactions being the most common approach [Kawahara 2013, Zheng 2019, Jie 2010]. Techniques to support paper engineering have also been proposed, facilitating the design of paper interfaces [Michelle 2015]. These explorations highlight the diverse range of methods and techniques available for building paper interfaces. Furthermore, magnets have been extensively utilized in tangible interfaces, enabling tactile feedback and the construction of physical interfaces [Bianchi 2013, Zheng 2018, Ogata 2018]. The versatility of magnets enables their combination with other materials, while magnetic powder, when combined with paper, offers an excellent means of providing physical actions while retaining the properties of paper [Ogata 2015].

While traditional interactions in paper-based interfaces include touch, pull, press, bend, and fold [Kawahara 2013, Zheng 2019, Jie 2010, 2015], the integration of magnets expands the possibilities to include rotate, swipe, toggle, and enriched haptic feedback [Kawahara 2013, Zheng 2018, Ogata 2018, Ogata 2015]. Building upon these prior works, we explore the integration of magnets with paper to extend the functionality of paper-based interactions.

## 3. MAGNET-EMBEDDED PAPER-BASED INTERACTION

By employing different paper structures and arrangements of magnets, we can achieve diverse interactions. This section introduces five interaction designs based on square paper and discusses their underlying mechanisms.

We used neodymium magnets throughout the development process. These magnets possess axial magnetization, with poles located at the flat ends, providing a strong magnetic force within a small footprint. We can adjust the movement distance of each interaction by using different sizes of magnets and cardboard pieces. For this work, we chose to use a rectangular magnet of 1x5x10mm, which has a strong magnetic force while maintaining a thin thickness. We also used a 2x2 inch sheet of card stock, a shape and size we chose based on the ease of material preparation: we could cut exactly 36 pieces from a common 12x12 inch cardstock paper. We primarily use 65 lb cover (177gsm) cardstock paper. Other paper thicknesses are possible, but a cardstock that is too thick (e.g., 100lb cover/271gsm cardstock) will weaken the force between the magnets and make folding difficult, while a paper that is too thin (e.g., 20lb/75gsm copy paper) will make the paper structure too weak to carry the magnets around. We begin by adding two sets of magnets, positioned in different directions, to the four corners of the paper (Fig 1).

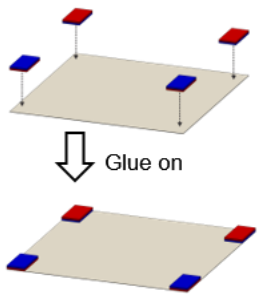

**Fig 1: Adding magnets to the square paper.**

As the two sets of magnets are oriented differently, folding the paper in distinct directions results in two different functional interactions. Folding the paper parallel to the short side of the magnet causes the two sets of magnets to attract each other. We attach the conductive tape to the paper, along with another set of magnets, transforming it into a power unit that can accommodate a cell battery (Fig 2). Folding the paper parallel to the long side of the magnets causes the two sets of magnets to repel each other, allowing for the creation of a momentary switch (Fig 3).

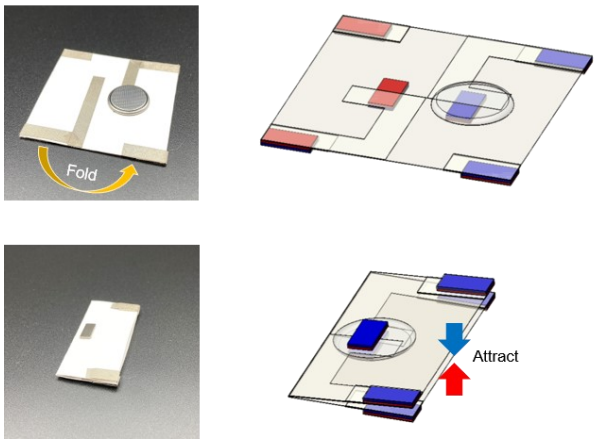

**Fig 2: Power Unit with a cell battery**

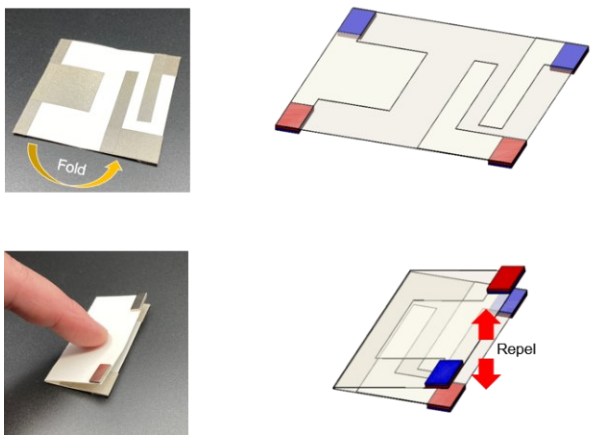

**Fig 3: Momentary Switch**

By leveraging the attractive and repulsive forces of different magnet faces, two types of alternating switches can be created by folding the paper multiple times (see Fig 4).

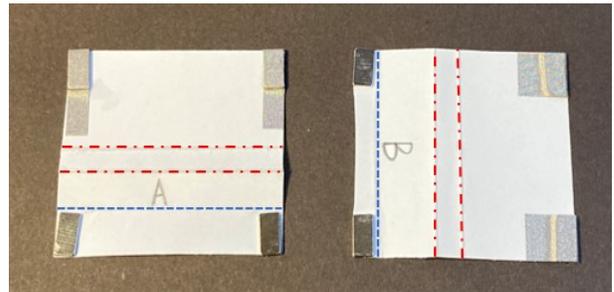

**Fig 4: Two alternating switches and their creases.**

Alternating switch A (Fig 5) is folded parallel to the short side of the magnets, causing the flat ends of the two pairs of magnets to attract each other. Pushing diagonally results in the top magnets' sides facing the sides of the bottom magnets, and when their sides repel each other, the top half of the card pops open.

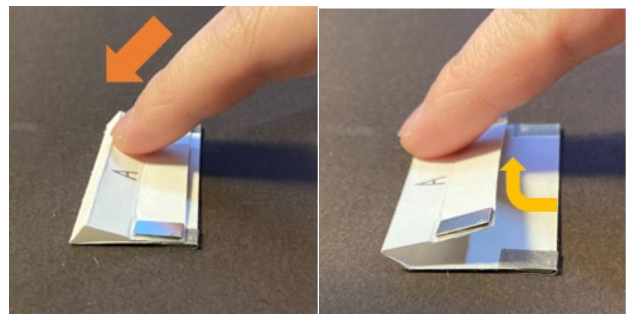

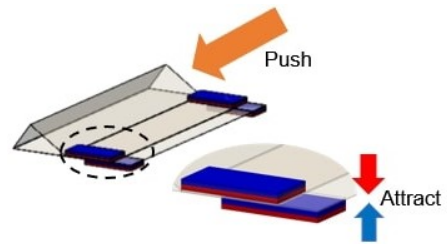

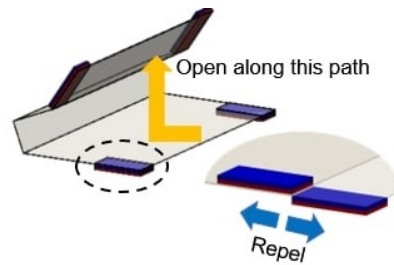

**Fig 5: Alternating Switch A**

Alternating switch B (Fig 6) is folded parallel to the long side of the magnets, causing the flat ends of the two pairs of magnets to repel each other, while their side ends attract each other. Pressing it from the top pushes the top magnets to align with and repel the magnets at the bottom.

| Type | Attraction **parallel** to the direction of magnetization | Repulsion **parallel** to the direction of magnetization | Attraction **perpendicular** to the direction of magnetization | Repulsion **perpendicular** to the magnetization direction |
|---|---|---|---|---|
| Diagram | 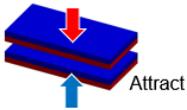 | 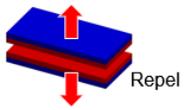 | 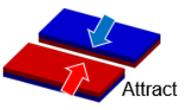 | 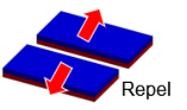 |
| Related Actions | squeeze, slide | pop up | slide | pop up |
| Example Units | 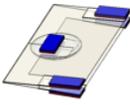 Power Unit | 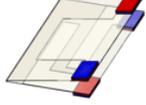 Momentary Switch | 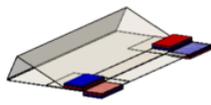 Alternating Switch B | 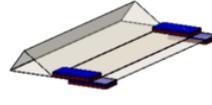 Alternating Switch A |
| | 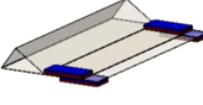 Alternating Switch A | 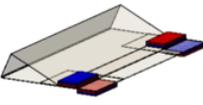 Alternating Switch B | 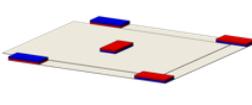 Slide Switch | |

**Table 1: Four types of magnetic forces and their related actions**

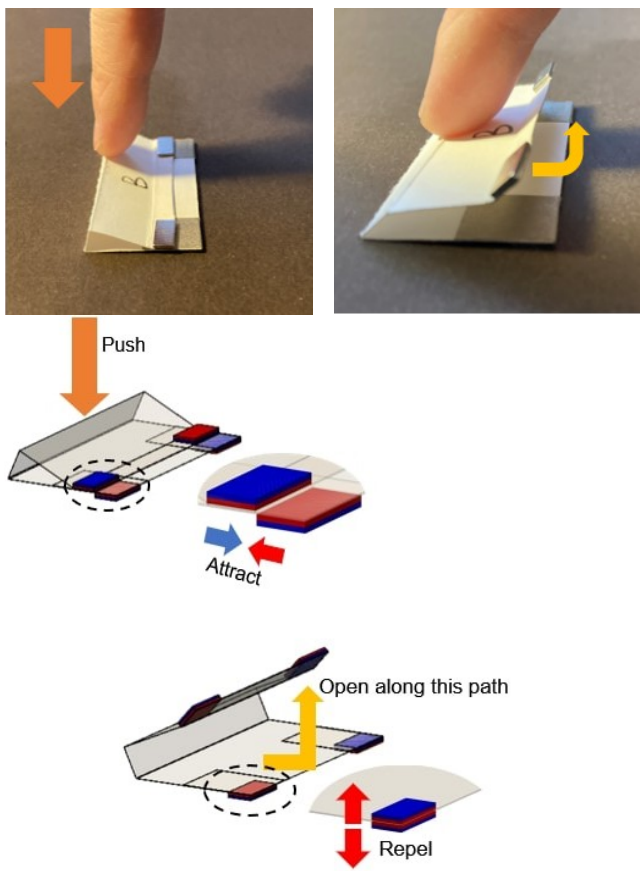

**Fig 6: Alternating Switch B**

The same mechanism is employed to create a slide switch (Fig 7) that utilizes the magnetic force from different sides of the magnet to facilitate the switching action.

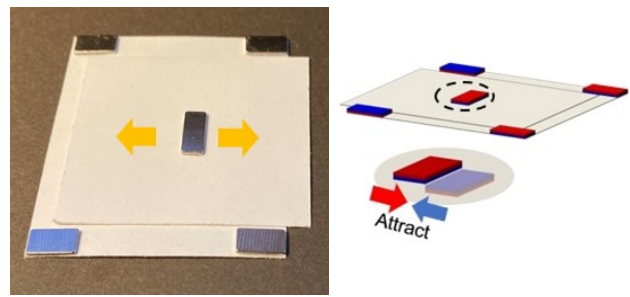

**Fig 7: Slide Switch that can slide left and right**

With the same magnet distribution, through the added folding and magnets, we can rapidly design and build paper-based interactions with different functions. These interactions are based on four types of magnetic forces under different magnet placements, and Table 1 summarizes these magnet forces and their related actions and representative units.

By introducing magnets into the design, we strike a balance between preserving the inherent flexibility and accessibility of paper while simultaneously reducing the need for intricate folding techniques or complex electronic components. This amalgamation of paper and magnets empowers users to create interactive experiences without the requirement for advanced technical expertise or extensive fabrication processes. The resulting system offers a user-friendly and versatile platform for designing paper-based interactions.

## 4. SOFT ROBOTICS AND BEYOND

The accessibility of paper-based interaction fabrication directly impacts the advancement of soft robotics. The integration of magnets onto paper simplifies the fabrication process, reducing reliance on specialized equipment and materials, and thereby making interaction design and prototyping more accessible to a broader audience. This fabrication approach enables researchers and hobbyists to iterate designs more efficiently, accelerating prototyping and development cycles. The proposed fabrication methods in this paper aim to foster innovation in the field of soft robotics.

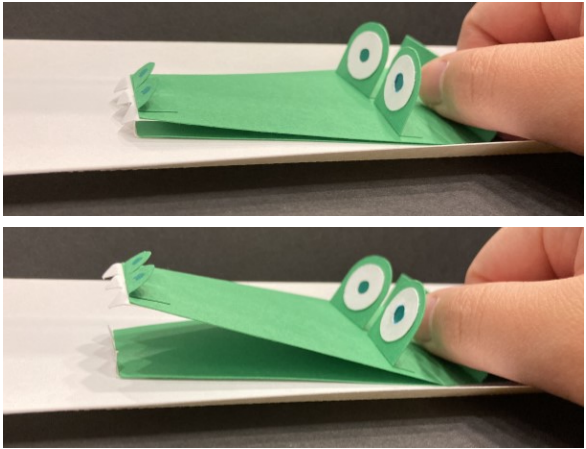

**Fig 8: Paper Crocodile**

As an illustrative example, we constructed a paper crocodile (Figure 8). By incorporating a magnet mechanism, when the paper crocodile is pushed, its mouth opens and closes. This movement adds a dynamic and responsive element to the otherwise static paper structure. The simplicity yet captivation of this behavior inspires curiosity and invites users to explore the boundaries between the physical and digital worlds. Furthermore, the paper crocodile serves as a catalyst for further discussion on potential applications and design considerations for paper-based interfaces in soft robotics. Exploring various mechanisms and magnet configurations could lead to the development of more complex and lifelike movements, thus enabling the creation of interactive paper objects with a heightened sense of realism.

Moreover, the impact of accessible paper-based interaction fabrication extends beyond soft robotics. The versatility of paper interfaces opens up possibilities for applications in wearable technology, interactive art installations, and educational projects. By lowering the barriers to entry, designers, artists, and educators can incorporate paper interfaces into their creations. This technique also promotes collaboration between different disciplines and enriches the intersection of technology, art, and education.

While paper and magnets offer unique advantages for rapid building and dynamic interaction, their practicality and suitability in a variety of situations need close consideration and continuous refinement. First, while paper can be used for simple structures and prototypes, its application becomes more challenging when dealing with complex designs or large projects. The structural integrity of complex paper structures may not be as reliable as those made of stronger materials, and too many magnets may interact with each other resulting in unstable and uncontrolled forces. In addition, when it comes to specific environments like water, paper proves to be an unsuitable choice because it is susceptible to moisture, which can cause it to deform or disintegrate. While magnets can improve the versatility and ease of building of paper, this approach still needs further exploration and development. The interaction between paper and magnets requires more in-depth study to determine its full potential.

## ACKNOWLEDGMENTS

This material is based upon work supported by the U.S. National Science Foundation under Grant No IIS-2040489.